\documentclass[useAMS]{mn2e}
\usepackage{color}
\usepackage{graphicx}
\usepackage{amsmath}
\usepackage{txfonts}

\title[Cyclotron resonance scattering feature in SMC~X-2]
{Detection of Cyclotron Resonance Scattering Feature in High Mass X-ray Binary Pulsar SMC~X-2}
\author[G. K. Jaisawal and S. Naik]{Gaurava K. Jaisawal\thanks{gaurava@prl.res.in} and Sachindra Naik\thanks{snaik@prl.res.in} \\
Astronomy and Astrophysics Division, Physical Research Laboratory, Navrangapura, Ahmedabad - 380009, Gujarat, India\\}

\begin{document}

\date{}

\maketitle

\begin{abstract}

We report broadband spectral properties of the high mass X-ray binary pulsar 
SMC~X-2 by using three simultaneous {\it NuSTAR} and {\it Swift}/XRT observations 
during its 2015 outburst. The pulsar was significantly bright, reaching a luminosity 
up to as high as $\sim$5.5$\times$10$^{38}$~ergs~s$^{-1}$ in 1-70 keV range. 
Spin period of the pulsar was estimated to be 2.37~s. Pulse profiles were found to be 
strongly luminosity dependent. The 1-70 keV energy spectrum of the pulsar was well 
described with three different continuum models such as (i) negative and positive power-law 
with exponential cutoff, (ii) Fermi-Dirac cutoff power-law and (iii) cutoff power-law models. 
Apart from the presence of an iron line at $\sim$6.4 keV, a model independent absorption like 
feature at $\sim$27~keV was detected in the pulsar spectrum. This feature 
was identified as a cyclotron absorption line and detected for the first time in 
this pulsar. Corresponding magnetic field of the neutron star was estimated to be 
$\sim$2.3$\times$10$^{12}$~G. The cyclotron line energy showed a marginal
negative dependence on the luminosity. The cyclotron line parameters were found to 
be variable with pulse phase and interpreted as due to the effect of emission 
geometry or complicated structure of the pulsar magnetic field.

\end{abstract}


\section{Introduction}

Cyclotron absorption lines are the unique features observed in the spectrum of 
accretion powered X-ray pulsars with magnetic field of the order of $\sim$10$^{12}$ G. 
These absorption like features are originated due to the resonant scattering between 
hard X-ray photons and electrons in quantized energy states  (M{\'e}sz{\'a}ros 1992).
These energy levels, known as Landau levels, are equi-spaced and depend on the strength 
of the magnetic field. The separation between Landau levels corresponds to the 
energy of the cyclotron resonance scattering feature and expressed by a relation 
{\it E$_{cyc}$=11.6B$_{12}\times(1+z){^{-1}}$ (keV)}, where  B$_{12}$ is the magnetic 
field in the unit of 10$^{12}$~G and $z$ is the gravitational red-shift. The detection 
of cyclotron absorption line is therefore an unique method to directly estimate the 
magnetic field of the accretion powered X-ray pulsars. Despite abundance of data from 
several previous X-ray missions, these features are only detected in $\sim$25 accretion 
powered X-ray pulsars  (Pottschmidt et al. 2012 and references therein).

SMC~X-2 is one of the brightest transient X-ray pulsar in the Small Magellanic Cloud (SMC). 
It was discovered with {\it SAS-3} in 1977 at a luminosity of 8.4$\times$10$^{37}$ ~ergs~s$^{-1}$ 
in 2-11 keV energy band, assuming a distance of 65 kpc (Clark et al. 1978; Clark, Li \& van 
Paradijs 1979). Since the discovery, the source was observed with several observatories such 
as {\it HEAO}, {\it Einstein} and {\it ROSAT}, which established the transient nature of 
the pulsar (Marshall et al. 1979; Seward \& Mitchell 1981; Kahabka \& Pietsch 1996).
Pulsation at $\sim$2.37~s  was discovered with the {\it RXTE} and {\it ASCA} observations
during one of the major X-ray outburst observed in 2000 (Torii et al. 2000; Corbet et al. 
2001; Yokogawa et al. 2001). The optical companion discovered by Crampton, Hutchings \& 
Cowley (1978) was later resolved into two early spectral type stars (Schmidtke, Cowley 
\& Udalski 2006). The I-band photometric studies of these stars with Optical Gravitational 
Lensing Experiment (OGLE) revealed that the northern star is the true optical companion of 
SMC~X-2. This conclusion was derived based on the periodic variability observed in the 
magnitude (up to 1 mag) at a period of 18.62$\pm$0.02 d (Schurch et al. 2011). A similar 
value of periodicity at 18.38$\pm$0.02 d was obtained from the pulse-period evolution 
studies of the pulsar from {\it RXTE} observations (Townsend et al. 2011). Observed 
periodicity of $\sim$18.4 d from two different approaches corresponds to orbital period 
of binary system. McBride et al. (2008) identified the optical companion as a O9.5 III-V 
emission star.

Since 2000, there was no report of any major X-ray activity (outburst) detected in the 
pulsar. Recently, an intense X-ray outburst was detected in 2015 September during which 
the pulsar luminosity reached up to as high as $\sim$10$^{38}$ ~ergs~s$^{-1}$ (Negoro et 
al. 2015; Kennea et al. 2015). The pulsar spectra obtained from {\it XMM}-Newton and 
{\it Swift}/XRT observations during this outburst were described with cutoff power-law 
model with a soft blackbody component at $\sim$0.15~keV. In addition to a hard spectrum 
($\Gamma$$\simeq$0), several emission lines from ionized N, O, Ne, Si and Fe were detected 
during this outburst (La Palombara et al. 2016). We have studied the 1-70 range energy 
spectrum of the pulsar by using the Nuclear Spectroscopy Telescope Array ({\it NuSTAR}) 
and {\it Swift}/XRT observations during the 2015 outburst. The details on observations, 
results and conclusions are presented in following sections of this letter.

\begin{figure}
\centering
\includegraphics[height=3.1in, width=2.2in, angle=-90]{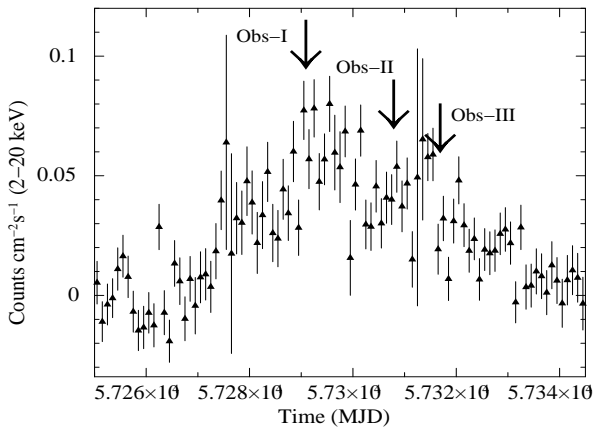}
\caption{Light curve of the 2015 outburst of SMC~X-2, observed with {\it MAXI} 
in 2-20 keV energy range covering the duration from 2015 August 16 (MJD 57250) 
to 2015 November 19 (MJD 57345), is shown. Arrows in the figure show the date of 
{\it NuSTAR} and {\it Swift}/XRT observations during the outburst.}
\label{maxi}
\end{figure}

\section{Observation and Analysis}

Following the report of X-ray outburst, Target of Opportunity (ToO) observations 
of SMC~X-2 were performed with {\it NuSTAR} (Harrison et al. 2013) at three different
epochs in 2015 September -- October as shown in Fig.~\ref{maxi}. 
Simultaneous {\it Swift}/XRT (Burrows et al. 2005) observations were 
also carried out at these epochs of the outburst. The details of the 
observations used in this paper are listed in Table~\ref{obs}. Hereafter, 
we used Obs-I (ObsIDs: 90102014002 \& 00034073002), Obs-II (ObsIDs: 90102014004 
\& 00081771002) and Obs-III (ObsIDs: 90101017002 \& 00034073042) to denote the 
first, second and third sets of {\it NuSTAR} and {\it Swift}/XRT data, respectively.   

{\it NuSTAR} is the first hard X-ray focusing telescope sensitive in the 3-79 keV 
energy range. It consists of two independent grazing incident telescopes that focus 
the photons at two different focal planes, FPMA and FPMB. We have used standard NUSTARDAS 
software v1.4.1 of HEASoft version 6.16 to generate the barycentric corrected light curves, 
spectra, response matrices and effective area files. The source light curves and spectra 
were extracted from the FPMA and FPMB data by selecting a circular region around the source 
center.  The radii of the circular regions used were 135, 100 and 100 arcsec for first, 
second and third {\it NuSTAR} observations, respectively. Background data products were 
accumulated in a similar manner by selecting circular regions of same size as quoted
above and away from the source. We also used the {\it Swift}/XRT data for spectral 
analysis below 10 keV. We used the standard \texttt{XRTPIPELINE} for reprocessing the 
XRT data. The source and background spectra were extracted from the window timing mode 
event by considering the source and background regions in XSELECT package. The response 
file for XRT was generated by using the \texttt{xrtmkarf} tool.


 \begin{table}
\centering
\caption{Log of simultaneous observations of SMC~X-2 with {\it NuSTAR} and {\it Swift}/XRT.}
\begin{tabular}{lccc}
\hline
\\
Observatory/		&ObsID   	    &Start Date   		&Exposure  \\
Instrument		   &	     	      &(MJD)			       &(ks) \\
\hline
{\it NuSTAR}		 &90102014002 	  &2015-09-25T21:51:08	  &24.5 \\ 
{\it Swift}/XRT  &00034073002   &2015-09-25T22:32:58    &1.8 \\
\\
{\it Swift}/XRT  &00081771002   &2015-10-12T21:30:58    &1.5 \\ 
{\it NuSTAR}		 &90102014004	  &2015-10-12T21:41:08	  &23 \\ 
\\
{\it Swift}/XRT  &00034073042   &2015-10-21T14:08:58    &4 \\
{\it NuSTAR}     &90101017002   &2015-10-21T21:31:08    &26.7\\
\hline
\end{tabular}
\label{obs}
\end{table}  


\section{Results}

\begin{figure}
\centering
\includegraphics[height=2.6in, width=2.8in, angle=-90]{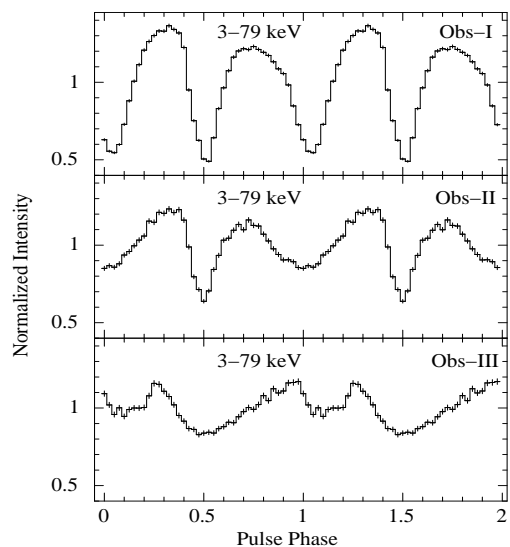}
\caption{Pulse profiles of SMC~X-2 obtained from the background subtracted light curves of 
FPMA detector of {\it NuSTAR} during first, second and third observations, respectively. Luminosity 
dependence of the pulse profiles can be clearly seen. The error bars in each panel represent 
1$\sigma$ uncertainties. Two pulses in each panel are shown for clarity.}
\label{pp}
\end{figure}

Source and background light curves from {\it NuSTAR} data were extracted at 
50 ms time resolution. We used the $\chi^2$-maximization method to estimate 
the barycentric corrected pulse period of the pulsar. The pulsation period at 
2.37197(2), 2.37141(2) and 2.37257(2) s were detected in the X-ray light curves
of the pulsar during Obs-I, Obs-II and Obs-III {\it NuSTAR} observations, 
respectively. The error in the pulse period was estimated for 1$\sigma$ 
significance level. The pulse profiles in 3-79 keV range were generated 
by folding  the background subtracted light curves at corresponding 
estimated spin periods from these observations and shown in Fig.~\ref{pp}. A 
strong luminosity dependence of the pulse profile can be clearly seen in the figure. During 
Obs-I (at the peak of the outburst), the pulse profile (top panel of Fig.~\ref{pp}) 
appeared double-peaked indicating the emissions or viewing of both the poles of the 
pulsar. However, significantly changed pulse profiles were observed during Obs-II 
\& III (at the declining phase of the X-ray outburst) and are shown in second and 
third panels of the figure.

\subsection{Pulse-phase-averaged spectroscopy}

Spectral properties of SMC~X-2 in broad energy range (1-70 keV)
have been investigated for the first time and reported in this
paper by using data from simultaneous observations with {\it NuSTAR} 
(3-70~keV range) and {\it Swift}/XRT (1-8~keV range). The procedure for 
spectral extraction was described in previous section. The source spectra 
were grouped to achieve $>$20 counts per channel bins. With appropriate 
background spectra, response matrices and effective area files, 
 simultaneous spectral fitting in 1-70 keV range was carried 
out by using {\it Swift}/XRT and {\it NuSTAR} data for each of the 
three epochs of observations. $XSPEC$ (ver. 12.8.2) package was used 
to do the spectral fitting. The spectral parameters were tied together 
during the fitting except the relative detector normalizations. 

\begin{figure}
\centering
\includegraphics[height=2.8in, width=3.6in, angle=-90]{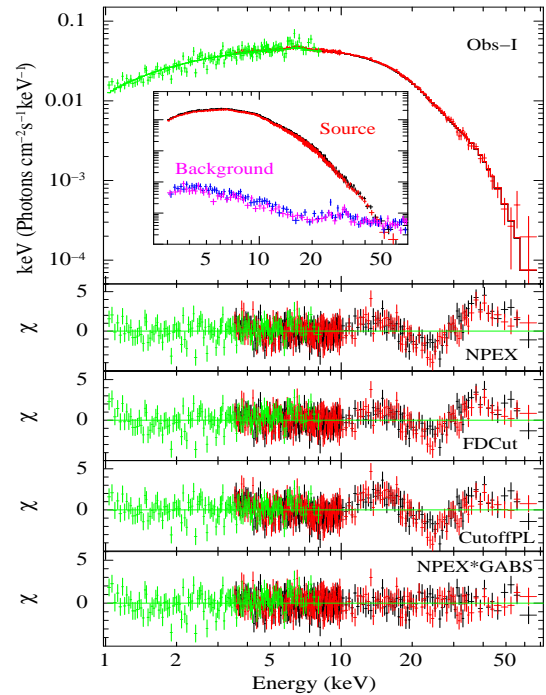}
\caption{The energy spectrum of SMC~X-2 in 1-70~keV range obtained from FPMA and 
FPMB detectors of {\it NuSTAR} and {\it Swift}/XRT data during the first observation 
along with the best-fit model comprising a NPEX continuum model with an iron emission 
line and a Gaussian absorption component for cyclotron resonance scattering feature. The 
second, third and fourth panels show the contributions of the residuals to $\chi^{2}$ when 
pulsar continuum was fitted with NPEX, FDCut and CuttoffPL models, respectively. In all 
these panels, an absorption like feature in 20-30 keV range is clearly visible.
The fifth panel shows the residuals for NPEX model after including a GABS component for 
cyclotron line. The background subtracted source spectra and background spectra 
are shown in the inset. It can be seen that the background is relatively much lower 
than the source counts during the observation.}   
\label{sp1}
\end{figure}

\begin{figure}
\centering
\includegraphics[height=2.8in, width=3.6in, angle=-90]{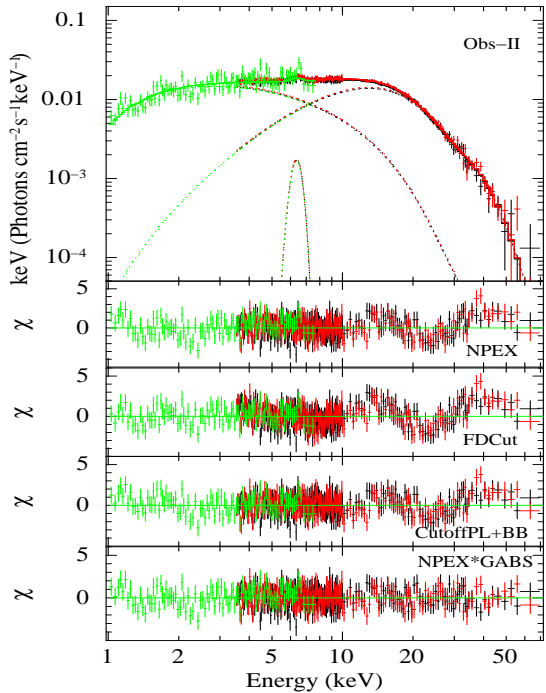}
\caption{The energy spectrum of SMC~X-2 in 1-70~keV range obtained from FPMA and FPMB detectors 
of {\it NuSTAR} and {\it Swift}/XRT data during second observation along with the best-fit model 
comprising a NPEX model with an iron emission line and a Gaussian absorption component for a 
cyclotron line. The second, third and fourth panels show the contributions of the residuals to 
$\chi^{2}$ when the continuum was fitted with NPEX, FDCut and CuttoffPL (with blackbody) models, respectively. 
In all these panels, an absorption like feature in 20-30 keV range is clearly visible.
The fifth panel shows the residuals for NPEX model after including a GABS component for 
cyclotron line.}   
\label{sp2}
\end{figure}

Standard continuum models such as high energy cutoff power-law (White et 
al. 1983), the Fermi-Dirac cutoff power-law (FDCut), the Negative and 
Positive power-law with Exponential cutoff (NPEX; Makishima et al. 1999) 
and cutoff power-law (CutoffPL) were used to fit the pulsar spectrum. 
We found that NPEX, FDCut and CutoffPL models can explain the pulsar 
continuum well for all three observations. However, an additional blackbody 
component was required to fit the source spectra obtained from Obs-II
\& Obs-III with CutoffPL model. An iron fluorescence emission line at 
$\sim$6.4~keV was detected in the source spectrum. Apart from the iron
emission line, an absorption-like feature at $\sim$27 keV was also 
detected in the pulsar spectra obtained from all {\it NuSTAR} observations. 
This feature was detected in the source continuum in a model independent 
manner. We added a Gaussian absorption line (GABS) in the model to explain 
the absorption feature. The addition of GABS in the model improved the
spectral fitting significantly yielding the reduced $\chi^2$ close to 1. 
The residuals obtained from the spectral fitting with all three models 
are shown in Fig.~\ref{sp1} \& \ref{sp2} for Obs-I \& II, respectively. 
A strong absorption like feature can be clearly visible in 20-30 keV 
energy range in second, third and fourth panels of the figures. 
This feature was found to be model independent and clearly detected 
in all three {\it NuSTAR} observations. We identified this feature as 
cyclotron absorption line of the pulsar. Best fitting parameters from 
the NPEX continuum model are given in Table~\ref{spec_par} for all three 
observations. It can be seen that the cyclotron line energies during these 
observations are marginally different, showing a negative dependence on 
luminosity. The source flux was found to be relatively high during Obs-I 
compared to that during Obs-II \& III. Although the cutoff energy is 
nearly same during these observations, the photon index is also showing 
a dependence on luminosity. A hard spectrum with photon index close to 
zero, i.e. almost flat spectrum, was observed during Obs-I.


\begin{table*}
\centering
\caption{Best-fitting parameters (with 90\% errors) obtained from the spectral fitting 
of three {\it NuSTAR} and {\it Swift}/XRT observations of SMC~X-2 during 2015 outburst 
with the NPEX continuum model with an iron emission line and cyclotron absorption line.}

\begin{tabular}{lcccccc}
\hline
Parameter                       &\multicolumn{2}{c}{Obs-I}       &\multicolumn{2}{c}{Obs-II}        &\multicolumn{2}{c}{Obs-III}    \\
                           &NPEX             &NPEX$\times$GABS   &NPEX       &NPEX$\times$GABS      &NPEX       &NPEX$\times$GABS  \\
\hline

N$_{H}$$^a$              &0.1$^{+0.03}_{-0.1}$  &0.18$\pm$0.06   &0.7$\pm$0.1    &0.5$\pm$0.1    &0.8$\pm$0.1     &0.6$\pm$0.1\\
Photon index              &-0.1$\pm$0.02         &0.04$\pm$0.04   &0.78$\pm$0.05  &0.62$\pm$0.08  &1.05$\pm$0.09   &0.81$\pm$0.09\\
E$_{cut}$ (keV)	          &4.4$\pm$0.1           &4.8$\pm$0.1     &4$\pm$0.1      &4.6$\pm$0.2    &4$\pm$0.1       &4.5$\pm$0.2  \\
Fe line energy (keV)             &6.42$\pm$0.08          &6.42$\pm$0.08   &6.32$\pm$0.09  &6.35$\pm$0.08  &6.31$\pm$0.15  &6.34$\pm$0.1   \\
Fe line eq. width (eV)           &47$\pm$10              &72$\pm$14       &88$\pm$18      &83$\pm$18      &70$\pm$20      &70$\pm$20\\
\\
Cycl. line energy (E$_{c}$) (keV)         &--               &27.2$\pm$0.9    &--             &29$\pm$1.6      &--              &29.8$^{+2.7}_{-1.7}$ \\
Cycl. line width ($\sigma_{c}$) (keV)     &--               &6.4$\pm$1.0     &--             &7$\pm$1.5       &--              &7.2$^{+2.3}_{-1.4}$ \\
Cycl. line strength ($\tau_{c}$)	            &--            &6$\pm$2         &--             &9$^{+5}_{-3}$   &--              &9$^{+9}_{-4}$ \\
\\
Luminosity$^b$ (1-70 keV)   &--                  &5.5$\pm$0.5     &--             &2.7$\pm$0.3      &--             &1.8$\pm$0.2       \\ 
Reduced-$\chi^2$ (dofs)     &1.55 (582)          &1.04 (579)      &1.29 (551)     &1.02 (548)      &1.27 (596)      &1.08 (593)  \\
\hline
\end{tabular}
\\
\flushleft
$^a$ : Equivalent hydrogen column density (in 10$^{22}$ atoms cm$^{-2}$); 
$^b$ : Luminosity in 10$^{38}$  ergs s$^{-1}$, assuming a distance of 61~kpc (Hilditch, Howarth \& Harries 2005). 
\\
\label{spec_par}
\end{table*}


\subsection{Pulse-phase-resolved spectroscopy}

Phase-resolved spectroscopy was performed to understand the variation of
the parameters of cyclotron resonance scattering features with the pulse
phase of the pulsar. As the pulsar was relatively bright during the first 
observation, we accumulated the phase-resolved spectra in 8 pulse-phase bins 
from the first {\it NuSTAR} observation. Using corresponding background, 
response and effective area files, the phase-resolved spectroscopy was carried 
out in the 3-70 keV energy range. The NPEX continuum model was used to fit 
the phase-resolved spectra. The absorption-like feature, as seen in phase-averaged 
spectra, was also clearly detected in each of the phase-resolved spectra. A GABS 
component was included in the continuum model for the cyclotron absorption line. 
While fitting, the  equivalent hydrogen column density (N$_{H}$), iron line 
parameters and cyclotron width were fixed to corresponding phase-averaged values 
as given in Table~\ref{spec_par}. 

\begin{figure}
\centering
\includegraphics[height=2.5in, width=2.8in, angle=-90]{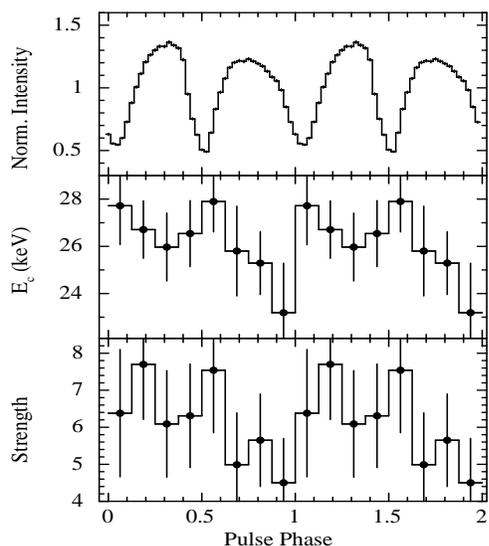}
\caption{Spectral parameters (with 90\% errors) obtained from the phase-resolved 
spectroscopy of SMC X-2 during first {\it NuSTAR} observation. Top panel shows the 
pulse profile in the 3-79 keV energy range. The values of cyclotron line parameter 
such as energy E$_c$ and strength $\tau$ are shown in second and third panels, respectively.}
\label{prs}
 \end{figure}

The cyclotron line parameters such as line energy and strength obtained from the 
phase-resolved spectroscopy are shown in Fig.~\ref{prs}. Both the parameters are
 marginally variable with pulse-phases of the pulsar. The energy of cyclotron 
line was found to be variable between 23 to 28 keV ($<$20\% of phase-averaged 
value) with maximum at around 0.5 pulse phase. The strength of cyclotron line 
was found to be varying between 4 to 8 and following similar pattern to that 
of cyclotron line energy. The errors in the figure are estimated for 90\% 
confidence level. We also checked the variation of cyclotron line energy 
with pulse phase by fixing the line strength at the phase-averaged value and 
vice versa. However, there was no notable change in the pulse phase dependence
of these parameters apart from a marginal improvement in the estimated errors
at each of the pulse phase values.

\section{Discussion and Conclusions}

As the magnetic field of accretion powered binary X-ray pulsars is in the order of 
10$^{12}$ G, the cyclotron resonance scattering features are expected to be detected 
in hard X-ray (10--100 keV) ranges. Therefore, the possibility of detection of such 
features in the hard X-ray spectrum becomes high when the source is very bright. 
During the 2015 giant X-ray outburst, the luminosity of SMC~X-2 was estimated to 
be significantly high ($\sim$5.5$\times$10$^{38}$~ergs s$^{-1}$) compared to earlier 
reported values. This high luminosity phase of SMC~X-2 enabled us to detect
the presence of cyclotron resonance scattering feature at $\sim$27 keV for the first time
in the pulsar spectrum which has not been reported earlier. As mentioned earlier, 
the cyclotron resonance scattering features are directly related to the magnetic 
field strength of the pulsar. Corresponding to the detected cyclotron line energy of 
$\sim$27 keV, the magnetic field of SMC~X-2 was estimated to be $\sim$2.3$\times$10$^{12}$ G. 
Although the cyclotron lines are rarely detected in the broad-band spectrum of accretion
powered binary X-ray pulsars, the hard X-ray focusing capability of {\it NuSTAR} 
remarkably contributed in the discovery of cyclotron lines in new sources. This 
helped in increase of the number of cyclotron sources to about 25.

During {\it NuSTAR} observations, the luminosity of the pulsar was very high ($>$10$^{38}$~ergs 
s$^{-1}$). It is possible that SMC~X-2 was accreting in the super-critical accretion regime during 
2015 X-ray outburst. At such high luminosity state, a radiation-dominated shock can be formed above 
the neutron star surface which decelerates infalling matter before settling on to the surface (Becker 
et al. 2007). Cyclotron lines are expected to be formed close to the shock region. As the luminosity 
increases, the shock region shifted upwards in the accretion column where relatively low value of
cyclotron line energy is observed. This results a negative dependence between cyclotron line energy 
and luminosity. We observed the expected negative correlation during {\it NuSTAR} observations of 
SMC X-2. This can be explained in terms of changes in shock height or line forming region with 
luminosity. Only two X-ray pulsars such as 4U~0115+63 (Nakajima et al. 2006) and V~0332+53 (Tsygankov 
et al. 2010) are known showing such negative correlation.  

For the first time, we presented a detailed phase-resolved spectroscopy of cyclotron 
parameters for SMC~X-2. The cyclotron line energy and its strength were found to be 
 marginally variable with pulse-phase of the pulsar. Numerical simulations based on cyclotron line 
features suggest that the 10-20\% variation in the cyclotron line parameters can be 
attributed to the viewing angle of the emission geometry. However, $>$30\% variation in 
parameters can be expected from distortion in magnetic dipole geometry of the pulsar 
(Sch{\"o}nherr et al. 2007; Mukherjee \& Bhattacharya 2012). During {\it NuSTAR} 
observations, the cyclotron line energy was variable within 20\% of the phase-averaged 
value which can be explained as the effect of viewing angle or local distortion in the 
magnetic field, as seen in other X-ray binary pulsars such as Cep~X-4 (Jaisawal \& Naik 2015) 
and GX~304-1 (Jaisawal et al. 2016). 

In summary, we report the discovery of cyclotron absorption line at $\sim$27 keV in SMC~X-2 
with {\it NuSTAR} and {\it Swift}/XRT observations at three epochs during 2015 X-ray outburst. 
The cyclotron line was detected in all three observations in a model independent manner. Using
the detected cyclotron line parameters, the magnetic field of the pulsar was estimated to be 
$\sim$2.3$\times$10$^{12}$ G. A negative dependence between cyclotron line energy and luminosity
in SMC~X-2 can be explained as due to the change in the shock height or line forming region with 
luminosity. The phase-resolved spectroscopy from first {\it NuSTAR} observation also revealed the 
presence of cyclotron line at different phases. The pulse-phase variation of the cyclotron parameters 
can be attributed as the effect of viewing angle or role of complicated magnetic field of the pulsar.


\section*{Acknowledgments}
We sincerely thank the referee for his/her valuable comments and suggestions which 
improved the paper. The research work at Physical Research Laboratory 
is funded by the Department of Space, Government of India. The authors would like to thank 
all the {\it NuSTAR} and {\it Swift} team members for ToO observations. This research has made 
use of data from HEASARC Online Service, and the NuSTAR Data Analysis Software (NuSTARDAS) 
jointly developed by the ASI Science Data Center (ASDC, Italy) and the California Institute of 
Technology (USA).

\end{document}